# UHD-DPDK Performance Analysis for Advanced Software Radio Communications

Daniel M. Brennan
Vuk Marojevic
dmb845|vm602@msstate.edu
Electrical and Computer Engineering
Mississippi State University
Mississippi State, MS

*Abstract*— Research conducted in LTE and 5G wireless communications systems uses common off-the-shelf hardware components and commercial software defined radio (SDR) hardware. One of the more popular SDR platforms is the Ettus USRP product line which uses the UHD driver and transport protocol framework. System performance can be increased using kernel bypass frameworks along with UHD. This paper investigates UHD with DPDK in an SDR environment using srsITE as the SDR application. We present measurement results using the iperf3 network performance application that show performance improvements when employing a kernel bypass framework to facilitate data transfer over the network interface between the SDR application and the radio hardware.

## I. INTRODUCTION

Advances in general-purpose computing hardware and open source software development efforts now enable research conducted in LTE and 5G wireless communications systems to be performed using common-off-the-shelf (COTS) x86_64 computers running Linux, open source software, and commercial software defined radio (SDR) hardware. One of the more popular SDR platforms is the Ettus USRP product line [1]. Ettus provides an open source driver set for interfacing with the USRP product line, known as the USRP Hardware Driver (UHD) [2]. Combined with multi-platform support, UHD provides an application programming interface (API) common to all USRP devices and portable for researchers using various SDR frameworks.

This research investigates the network interface between the host computer and the radio front end, or SDR hardware, and compare alternative access mechanisms, specifically Linux kernel-based versus the Data Plane Development Kit (DPDK) developed by Intel for their network interface cards (NICs).

In particular, this paper explores the performance improvements provided by UHD-DPDK with srsLTE [1] as the end-to-end SDR application. In this configuration srsLTE provides LTE system, consisting of the LTE core network, or Evolved Packet Core (EPC), the LTE base station, or eNodeB (eNB), and the user equipment (UE). For a single radio access network link that this analysis requires, two physical components using two computers with the srsLTE software, and two USRPs as used.

There has been increasing interest to use these open source software tools that implement advanced wireless protocols with SDR hardware for supporting experiments outside the laboratory. In the US, a major effort is currently ongoing to deploy three major testbeds for advances wireless research, known as Platforms for Advanced Wireless Research (PAWR). This research aims at supporting these platforms, specifically the Aerial Experimentation and Research Platform for Advanced Wireless (AERPAW) [1], and the researchers using them. The present analysis is applicable to SDRs using UHD and an Intel NIC to transfer samples between the radio device and the host computer, where the latter is doing most of the baseband and higher layer processing of signals and protocols.

## II. BACKGROUND

### A. UHD

Ettus Research, a National Instruments company, develops the Universal Software Radio Peripherals (USRPs), the widely used SDR hardware, and develops and maintains UHD, which provides a streaming architecture to facilitate the transfer of SDR samples between the user application and the USRP hardware. Fig. 1 is a simplified illustration of the UHD streaming architecture in UHD 3.14.

Note that certain USRP models interface with the host computer via USB 3.0 and also use UHD to facilitate the data flow. Those are the B-series USRPs from Ettus Research/National Instruments and are not the subject of this study. This study focused on the X3xx and N3xx series USRPs using 10 Gbe fiber interfaces.

The Linux system introduces jitter due to scheduler and other events which can be mitigated by utilizing larger hardware buffers. The drawback to making larger buffers is additional hardware complexity and cost. Also, this results in increased packet latency.

For low throughput applications, the latency and jitter introduced by the Linux network stack for data plane traffic is acceptable. However, for higher speed links the latency and jitter impairments can cause dropped, late, or missed packets. This reduces the effective link bit rate and, in some cases, prevents link establishment.

As shown in Fig. 1, UHD uses the host operating system's network stack to facilitate data transfer between the computer and USRP hardware. Transmit data flow in an SDR system employing UHD streaming is shown in Fig. 2. The receive path is similar.

The user application creates SDR samples and sends to the UHD streamer using a UHD API send() function, illustrated in Fig. 2. The UHD streamer converts the samples to UHD format and copies to an internal UHD streamer buffer. When UHD is ready to send, it packetizes the data and sends to the kernel socket interface using the boost.asio library [5], where the data is copied again and sent to the network driver.

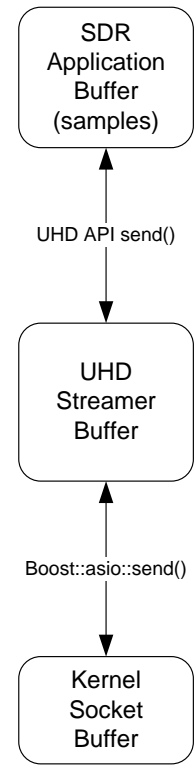

Fig. 2. Data flow in the UHD streaming architecture

The UHD streamer uses a credit-based flow control scheme. In this mode, the UHD streamer resizes the downstream buffers on the USRP to adjust the transmit window. When ready to transmit, UHD fills an internal transmit buffer, then starts sending packets to the network interface until the window is full. It then waits for a flow control response packet which indicates when free space is available. The credit-based flow control scheme is shown in Fig. 3.

To achieve the high bandwidth required for high-speed links several techniques can be employed to minimize latency and jitter. These include:
- Linux scheduler bypassing
- Minimized context switching between tasks
- CPU affinity
- User space drivers
- Hugepages

The Linux scheduler assigns threads to CPU cores when they are started. A thread may switch from the running state either voluntarily (for example, waiting for I/O) or involuntarily (pre-empted). In any case that thread will have to be rescheduled on the ready queue. The default Linux scheduler is the Completely Fair Scheduler (CFS), which was designed to share CPU cores fairly. CFS uses a time-ordered red-black tree to build a timeline of future task execution [6]. This is suboptimal in dedicated networking machines with significant network I/O and can introduce latency and jitter in network packet processing.

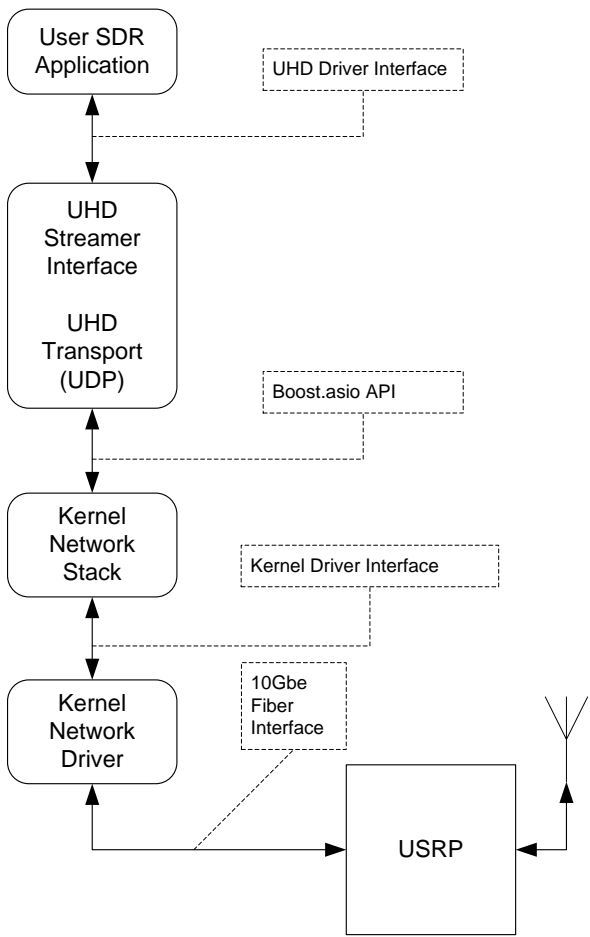

Fig. 1. UHD streaming architecture.

Context switches in the scheduler incur additional latency and jitter for threads. This can be minimized by isolating CPU cores and using CPU affinity to assign high priority networking threads to particular CPU cores.

The use of user space drivers removes the additional overhead of switching from user space to kernel space to facilitate the transfer of packets from the NIC to user buffers.

Finally, the use of hugepages to minimize page faults and keep critical code and data in memory without accesses to require secondary storage.

When used together these techniques can form a bypass of kernel mode code, including the networking stack, and offerhigher performance networking with lower latency.

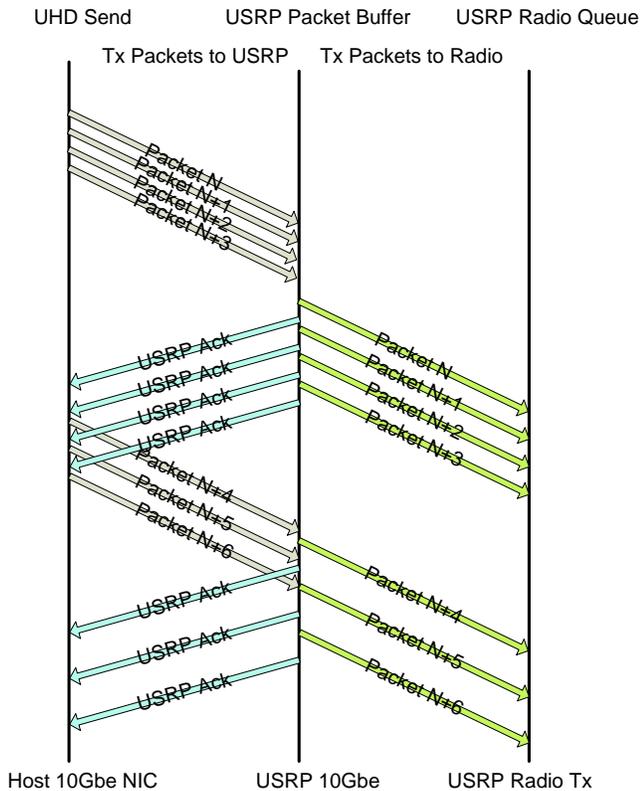

Fig. 3. UHD Streamer credit-based flow control.

## B. UHD-DPDK

Kernel bypass frameworks have been developed which circumvent the Linux networking stack and provide direct access from user space to the buffers and registers in the networking path. One such framework is the Data Plane Development Kit (DPDK) [7]. DPDK has been used in a wide variety of networking applications to provide a high-performance low-latency/low-jitter network interface. DPDK provides a well-documented API with numerous examples for users to port existing sockets-based applications to the DPDK framework.

Starting with version 3.14, the UHD driver includes a DPDK mode of operation for data plane traffic for networked USRPs. In this mode UHD replaces the boost.asio datagram socket interface with a DPDK interface to implement a true zero copy kernel bypass data path as shown in Fig. 4.

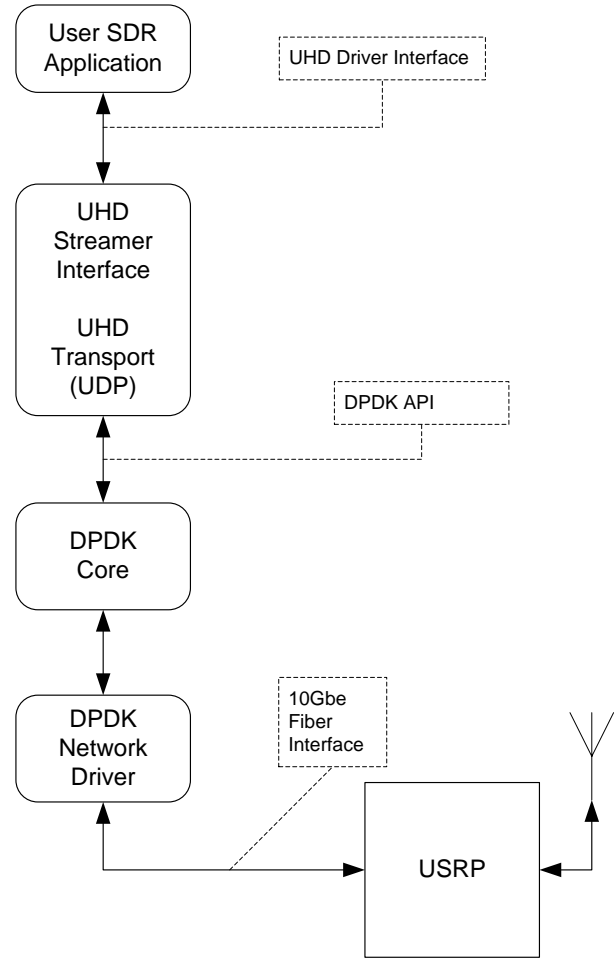

Fig. 4. Data flow in the UHD-DPDK streaming architecture

## III. RELATED WORK

Performance assessment of open source SDR software using srsLTE with UHD and USRP hardware measured throughput and CPU usage in [8]. This study compared srsLTE to another open source SDR framework, Open Air Interface.

Performance analysis using UHD-DPDK and GNU Radio for the SDR application was presented at the GNU Radio Conference 2019, Huntsville, AL [9]. The results indicate an improvement in both throughput and latency using UHD-DPDK for the underlying driver and data plane transport protocol. It was reported that latency improved approximately 73% using UHD-DPDK. The report claimed throughput improvements, but not significant enough due to bottlenecks in GNU Radio.

## IV. TEST AND MEASUREMENT CONFIGURATION

The hardware setup for this test and measurement includes two Intel x86_64 computers, one each for eNB/EPC and UE endpoints. A 10Gbe Intel NIC was used on the EPC/eNB computer and a USB 3.0 interface on the UE computer. The radio hardware employed two Ettus USRPs.

### A. Hardware Setup

The test configuration uses two computers from Supermicro Systems, Inc. Both are equipped with the X10SRW-F motherboards. The CPU used in each is an Intel(R) Xeon(R) CPU E5-1660 v3 @ 3.00GHz with a C610/X99 series chipset.

Each system has 32 GB RAM. The memory devices used in both systems are 16 GiB DDR4 Synchronous 2133 MHz (0.5 ns) from Micron (part number 36ASF2G72PZ-2G1B1).

The computer designated to the eNB and EPC has an additional Intel X520 82599ES-based 10-Gigabit SFI/SFP+ Network Interface Card to provide a 10 Gbe interface to the X310 USRP, which has two SFP+ interfaces. Only one interface was used in this experiment, which was populated with an Intel dual rate 10GBASE-SR/1000BASE-SX 850 nm multi-mode fiber optic transceiver (part number AFBR-703SDZ-IN2). The X520 network port was connected via a 5-meter multi-mode fiber optic cable to the USRP X310's Port 1, using the same model SFP+ transceiver.

The radio hardware consists of two Ettus USRPs; an X310 with a 10 Gbe interface connected to the eNB/EPC machine and a B210 with a 3.0 USB interface to the host computer taking the role of the UE.

The USRPs are direct connected via coaxial cable and a 30 dB attenuator. There is no wireless link used in this experiment to eliminate radio bottlenecks and ensure reproducible experiments.

The hardware setup is shown in Fig 5.

### B. Software Configuration

#### 1) Operating System
The systems run Ubuntu 18.04.1 server version. There are no virtual machines used in this test.

#### 2) UHD, DPDK, and srsLTE
The SDR system in this experiment requires 3 main components: hardware driver and user plane transport protocol (UHD), kernel bypass framework (DPDK), and SDR application (srsLTE). TABLE 1 lists the versions of these components and any other libraries or tools used to build the software.

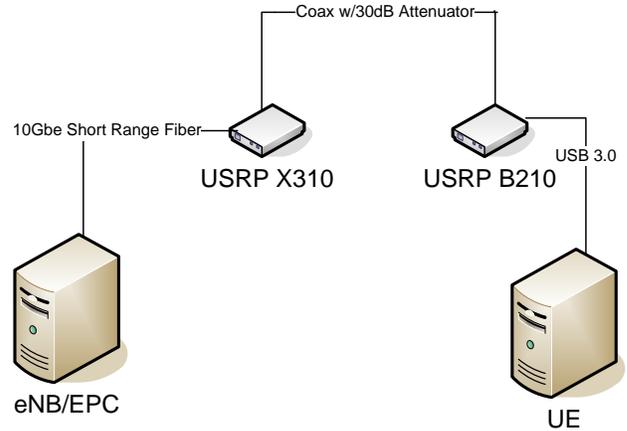

Fig. 5. Test configuration hardware setup

#### 3) srsLTE Configuration
The srsLTE applications have individual configuration files to set various run-time parameters. The eNB file is located in ~/.config/srsLTE/enb.conf.

The srsLTE eNB configuration was modified to set the transmit and receive gains as follows:

tx_gain = 80
rx_gain = 40

All other parameters in the enb.conf file were default settings except for the device_args setting. When running in DPDK mode the enb.conf file had the additional line:

device_args = use_dpdk=1,type=x300,addr=192.168.40.1

where 192.168.40.1 is the USRP hardcoded IP address for Port 1.

### C. Building the Code

All code was downloaded from repositories and compiled directly on the target test machines. Basic build configurations were used with modification to facilitate UHD-DPDK.

UHD was compiled using the -DENABLE_DPDK=ON to enable DPDK. This was done by specifying the switch in the cmake command in the build directory:

cmake ../ -DENABLE_DPDK=ON

By default, DPDK builds static DPDK libraries. However, shared libraries are required for UHD. Therefore, the DPDK build configuration file had to be modified to compile for shared libraries. This was done by modifying the

CONFIG_RTE_BUILD-SHARED_LIB line in the dpdk .config file to

CONFIG_RTE_BUILD_SHARED_LIB=y

TABLE 1. Key Software Components

| Component | Version | Description |
|---|---|---|
| Linux distribution | Ubuntu 18.04 | Operating system on host computer |
| Linux kernel | 5.3.0-28 | Linux kernel version |
| UHD | 3.15.LTS | Ettus UHD driver and transport protocol |
| DPDK | 17.11.9 | Data Plane Development Kit kernel bypass framework |
| srsLTE | 19.12 | Software Radio Systems LTE SDR framework, including eNodeB, EPC, and UE applications |
| gcc | 7.5.0 | GNU C compiler |
| glibc | 2.27 | GNU C library |
| iperf3 | 3.7+ | IP packet generator and performance analysis tool |

## V. TEST METHODOLOGY

The tests were selected to measure two components of UHD and UHD-DPDK: 1) end-to-end performance including throughput, jitter, and packet loss and 2) Linux CPU utilization of srsLTE and UHD/UHD-DPDK threads.

### A. End-to-End Performance

The iperf3 tool was used to measure the end-to-end throughput, jitter, and packet loss between the UE and EPC over the UE/eNB hardwired link. The iperf3 program was run in UDP mode to provide lost packet counters for each test. Test durations were 3 hours when possible to collect sufficient data to measure jitter and dropped packets. Connection issues, discussed below, prevented some tests from running long durations.

The eNB ran iperf3 in server mode listening for connections from the UE.

iperf -s

The UE ran iperf3 as a client, requesting the eNodeB transmit UDP data using the following command

iperf3 -u -b <bit_rate> -c 172.16.0.1 -R

where -b <bit_rate> specifies the rate at which the server side will transmit to the client. The selected bit rates we chosen so throughput, jitter, and drop measurements could be compared for both UHD and UHD-DPDK modes. These tests correspond to downlink throughput tests for a single UE attached to the LTE network. We consider 5, 10, and 20 MHz LTE system configurations, in frequency division duplex (FDD) mode, where the entire 20 MHz radio access network channel can be scheduled for the uplink or downlink, or both, to a single user.

### B. Thread CPU Utilization

The Linux top [10] command was used to capture CPU utilization of the processes and threads related to UHD, DPDK and srsLTE. Linux process monitoring was performed on the eNB only since that is the machine of interest comparing UHD to UHD-DPDK.

## VI. TEST RESULTS

### A. Performance Test Results

Tests were performed for different Physical Resource Blocks (PRB), corresponding to different LTE system bandwidths, and offered data rates, as shown in Table 2. Tests indicate UHD-DPDK a performance improvement in jitter and dropped packer percentage over UHD. Note that UHD could not establish a connection when using 100 PRB, or 20 MHz LTE. In addition, even though UHD-DPDK did connect and transmitted data, the performance was worse than for 50 or 25 PRB in terms of dropped packets. It is our experience in the lab setting that the B210/X310 pair do not perform well using UHD and often have a difficult time connecting at PRB above 50. Connection establishment was near 100% using UHD-DPDK at any sampling rate.

As the results indicate, UHD-DPDK improves jitter and packet drop percentage for any configuration that we tested.

TABLE 2. Jitter and dropped packet test results (N/C – does not connect)

| | | UHD | | UHD-DPDK | |
|---|---|---|---|---|---|
| PRB | Data Rate (Mbps) | Jitter (ms) | Drop % | Jitter (ms) | Drop % |
| 25 | 10 | 0.669 | 0.0072 | 0.518 | 0.0056 |
| 50 | 30 | 0.448 | 0.23 | 0.298 | 0.018 |
| 100 | 30 | N/C | N/C | 0.179 | 0.11 |

### B. Thread CPU Utilization Results

The 'top' command was run in batch mode for each test iteration, producing average CPU utilization every 4 seconds for threads used by both srsLTE, UHD and UHD-DPDK. Only threads involved in the srsLTE PHY module and those that produced any appreciable CPU percent utilization (> 1) are shown in Table 3.

The default Linux scheduler is running on the system, with both srsLTE and UHD running their individual threads with a real-time priority. Neither the srsLTE eNB (srsenb) application nor UHD use CPU affinity to assign threads to CPU cores.

Running srsLTE with UHD it is observed that work is mainly scheduled among the same 3 CPU cores during each. The percent CPU utilization is shown in Table 3.

When using DPDK and additional core is required since DPDK assigns each DPDK I/O thread to its own core, thus typically consuming 4 CPU cores. UHD-DPDK uses CPU affinity to assign the DPDK I/O thread to a CPU core. Note that hyperthreading is disabled for all tests.

It is evident from Table 3 that CPU utilization is lower for srsLTE threads WORKER0-2, TXRX, ENBSOCKETS, STACK, and PRACH_WORKER in UHD-DPDK mode. This is due to kernel bypass provided by DPDK, freeing application threads to perform more PHY or protocol processing.

The UHD driver uses the boost asio library to facilitate transfers to and from the kernel networking stack. UHD employs the thread named "zero_copy_recv" to offload the main threads by handling packet reception with the boost.asio interface. This thread is not started in UHD-DPDK. On the other hand, dpdk-io thread is launched that handles the data transfer directly from the corresponding worker thread and the NIC.

The DPDK framework uses CPU affinity to assign execution units (threads) to specific cores. The UHD-DPDK implementation creates a DPDK thread for each receive port. For this test we used a single 10 Gbe port, thus a single dpdk-io thread was created. Note that DPDK threads in general utilize 100% of the core they run on. Therefore, we can see that dpdk-io averages 99.84% CPU utilization. srsLTE also uses CPU affinity and the worker threads can run on different cores to maximize performance.

TABLE 3. Linux process average CPU percent utilization

| Thread Name | UHD | UHD-DPDK |
|---|---|---|
| WORKER0 | 30.04 | 26.69 |
| WORKER1 | 31.29 | 27.80 |
| WORKER2 | 0.20 | 0.056 |
| TXRX | 10.34 | 9.65 |
| ENBSOCKETS | 6.14 | 5.97 |
| STACK | 7.42 | 6.52 |
| PRACH_WORKER | 2.47 | 2.36 |
| zero_copy_recv | 11.24 | N/A |
| dpdk-io | N/A | 99.84 |

*C. Radio Link Establishment and Connection Observations*

Though not directly tested, we noticed that the B210 and X310 USRPs do not interoperate very well at high sampling rates (i.e., above 50 PRB). We failed to make connect the radio link at the higher rates using UHD alone. Even though UHD-DPDK did connect, it took much longer to establish the radio link, and often times required several retries to establish the link. The process of link establishment can deal with different hardware using different oscillators which need to be aligned to account for frequency offsets. The LTE protocol provides this feature as part of the initial cell search through the Primary and Secondary Synchronization Signals.

We also observed link stability issues using UHD with the B210 and X310, with the link disconnecting at various times. This was rarely observed with UHD-DPDK.

The issue is most likely due to higher jitter and packet loss at the higher rates, which are evident in UHD-DPDK. The UE-to-network attachment process evaluates the quality of the radio channel in the broad sense, which includes the RF impairments. If a UE cannot properly synchronize or maintain a block error rate below 10 %, even after adaptive modulation and coding to te most robust transmission mode, the UE will disconnect and search for other cells.

VII. CONCLUSIONS

This paper reports a performance assessment of USRP hardware and srsLTE SDR application using the DPDK kernel bypass framework. Based on the results from our tests we conclude that a kernel bypass framework such as DPDK improves link performance and stability in an SDR environment.

Additional performance gains can be achieved by using CPU affinity in the SDR application and UHD driver on top of DPDK. This requires machines with high core count. The number of cores increases with each SDR application and network interface running on the machine. It is recommended to use Xeon-class machines with high performance network interfaces supporting advance Intel I/O acceleration technologies [11].

Porting applications to bypass the kernel stack with DPDK is straightforward as it is agnostic to the application framework. Following the documentation and examples found in [4] existing applications can be ported to increase network I/O performance, thus freeing up cycles for application processing. Further studies should explore DPDK in other areas, such as 5G fronthaul and backhaul.